\documentclass[12pt,preprint]{aastex}

\shorttitle{Minivoids in the Local Volume}
\shortauthors{Tikhonov A.V. and Karachentsev I.D.}

\begin{document}
\title{Minivoids in the Local Volume}
\author{Anton V.\ Tikhonov}
\affil{ St. Petersburg State University, St.\ Petersburg, Russia}
\email{avt@gong.astro.spbu.ru}
\author{Igor D.\ Karachentsev}
\affil{Special Astrophysical Observatory, Russian Academy
  of Sciences, N.\ Arkhyz, KChR, 369167, Russia}
\email{ikar@luna.sao.ru}
\begin{abstract}

 We consider a sphere of 7.5 Mpc radius, which contains 355 galaxies with
accurately measured distances, to detect the nearest empty
volumes. Using a simple void detection algorithm, we found six
large (mini)voids in Aquila, Eridanus, Leo, Vela, Cepheus and
Octans, each of more than 30 Mpc$^3$. Besides them, 24 middle-size
"bubbles" of more than 5 Mpc$^3$ volume are detected, as well as
52 small "pores". The six largest minivoids occupy 58\% of the
considered volume. Addition of the bubbles and pores to them
increases the total empty volume up to 75\% and 81\%,
respectively. The detected local voids look like oblong potatoes
with typical axial ratios b/a = 0.75 and c/a = 0.62 (in the
triaxial ellipsoide approximation). Being arranged by the size of
their volume, local voids follow power law of volumes-rankes
dependence. A correlation Gamma-function of the Local Volume
galaxies follows a power low with a formally calculated fractal
dimension D = 1.5. We found that galaxies surrounding the local
minivoids do not differ significantly from other nearby galaxies
on their luminosity, but have appreciably higher hydrogen
mass-to-luminosity ratio and also higher star formation rate. We
recognize an effect of local expansion of typical minivoid to be
$\Delta H/H_0\sim (25\pm15)$\%

\end{abstract}

\section{Introduction}

With the discovery of the first cosmic voids in the constellation
Bootis (Kirshner et al., 1981) and in other sky regions (Joveer et
al., 1978, Gregory and Tompson, 1978, de Lapparent et al., 1986,
Rood, 1988), a new branch of extragalactic astronomy ---
``voidology'' was formed. The concept of voids as basic elements
of large-scale structure of the Universe proved to be required in
modern cosmological models. Zeldovich et al., 1982 considered
giant volumes between superclusters which are almost empty of
visible objects in the framework of structure formation theories.
Ghigna et al. (1994) found that CDM models produce smaller
void-probability function (VPF) then the observational one
calculated on the volume-limited sub-sample of the Perseus-Pisces
survey. Sheth and van de Weygaert (2004) presented a model for the
distribution of void sizes and it's evolution in the context of
hierarchical scenarios of gravitational structure formation.
Furlanetto and Piran (2005) produced a straightforward analytical
model of the void distribution and of galaxy populations within
voids. Patiri et al. (2005) compared statistical analysis of voids
in 2dFGRS with that of cosmological $N$-body simulations and found
that galaxies in voids are not randomly distributed but form
structures like filaments. Icke (1984) on the basis of numerical
studies concluded that voids beetwen filamets are likely to be
roughly spherical. Regoes and Geller (1991) found that in their
models of structure formation in two and three dimensions,
suitable initial conditions lead to cellular structure with
faceted voids similar to those observed in redshift surveys. Icke
and Van de Weygaert (1991), van de Weygaert (1994) introduced
geometrical model - Voronoi foam (tessellations) for the skeleton
of cosmic mass on scales $10 - 200$~Mpc that includes voids as
building blocks of large scale structure. Sahni et al. (1994)
applied adhesion approximation to study the formation and
evolution of voids in the universe. Peebles (2001) gave detailed
review of modern status in theory and observations of voids.

Cosmic voids are usually considered to have typical dimensions of
$\sim$(20--50) Mpc, but the depth of voids, $\delta\rho/<\rho>$,
is the parameter which is still in question. Numerical modeling of
the large-scale structure in $\Lambda$CDM models (Van de Weygaert
and van Kampen, 1993, Gottlober et al., 2003, Sheth et al., 2004,
Colberg et al., 2005) reproduces cosmic voids of different sizes
and density contrast without conflicting with their typical
observational parameters.

It is evident that the most reliable characteristics of voids can
be obtained by investigating the nearest ones. In his Catalog and
Atlas of Nearby Galaxies, Tully (1988) noted the presence in the
Local Supercluster (LSC) of the so-called Local void which begins
directly from the boundaries of the Local Group and extends in the
direction of North Pole of the LSC by $\sim$20 Mpc. The Local void
looks practically free from galaxies. Karachentseva et al. (1998)
undertook a search for new dwarf galaxies in the region of Tully
void, but failed to find in it dwarf objects with luminosities
down to $10^{-3}$ of the luminosity of the Milky Way. Recently
Iwata et al. (2005) have measured velocities and distances of
galaxies in a vicinity of the Local void and derived some evidence
of its expansion. Studying the distribution of nearby galaxies in
a volume of radius 10 Mpc, Karachentsev (1994) noted the presence
in it of small voids of different sizes completely free from
galaxies. As new galaxies were discovered in this Local Volume and
their individual distances were specified, the foamlike pattern
structure of the Local Volume became more and more clear. This
feature may prove to be universal, being a fractal continuation of
the spectra of voids to small scales. At the present time, the
sample of galaxies with distances less than 10 Mpc numbers about
450 galaxies. For half of them the distances have been measured to
an accuracy as high as 8--10\% (Karachentsev et al., 2004=CNG).
The study of properties of voids in the most nearby sample,
restricted by the distance, has an obvious advantage since we see
here dwarf systems down to their minimum size/luminosity. This
gives us unique possibility to define voids as regions empty of
any type galaxies. According to scenarios of hierarchically
evolving structure in the Universe voids are expected to contain
some mass. But in this context our definition seems not to be
unphysical because inside our voids we have certain amount of
haloes of low masses that don't contain galaxies for different
reasons. The absence of the effect of ``God's fingers'' in the
Local Volume because of the virial motions of galaxies simplifies
the analysis of the shape and orientation of nearby voids.

Below we undertake the first attempt to map voids in the Local Volume on
the basis of quantitative reception of their finding.
Here we present a list of nearby minivoids, maps of their
distribution over the sky, and also discuss their shapes,
orientations and characteristics of galaxies neighboring with voids.

\section{The sample used for the void analysis}

Until 2000, very little data had been available to describe the
true 3D distribution of nearby galaxies even just around the Local
Group. This surprising situation has been overcome recently with
accurate distance measurements of nearby galaxies based on the
luminosity of the tip of the red giant branch (TRGB). This method
has a precision comparable to the Cepheid method, but requires
much less observing time. Over the last five years, snapshot
surveys with HST have provided us with the TRGB distances for many
nearby galaxies obtained with an accuracy of 8 - 10\%.

Apart from 36 members of the Local Group, there are so far 355
galaxies with distance estimates less than 7.5 Mpc. Among them,
accurate distances to 183 galaxies have been measured based on the
Cepheid method (N=12), or the luminosity of the TRGB (N=171). The
remaining galaxies have only rough (25 - 30\%) distance estimates
from the luminosity of their brightest stars, the Tully-Fisher
relation, or from their apparent membership to the known nearby
groups. In this restricted distance range the TRGB method is most
effective because it provides accurate distances to galaxies of
all morphological types with minimal observational demands. A list
of these galaxies is presented in "Catalog of Neighboring
Galaxies" by Karachentsev et al. (2004). The distribution of this
galaxy sample over the sky is shown in Fig.1 by filled circles of
different sizes according to distances of galaxies.

\section{Void detection algorithm}
There is no commonly accepted definition of void in galaxy
distribution still. Different algorithms of void-finding give
different list of voids. Some of them predefine shape of void
(e.g. spherical), some search for volumes of arbitrary shape. It
is worth to note the empty sphere method (Einasto, Einasto and
Gramann, 1989), the variant with elliptical volumes (Ryden and
Melott, 1996), the progressive construction of voids with cubes +
rectangular prisms (Kaufmann and Fairall, 1991),  the related
method that uses connected spheres (El-Ad and Piran, 1997), the
method of distance field maxima (Aikio and Mahonen, 1998), method
of connected underdense grid-cells (Plionis and Basilakos, 2002),
the use of the smoothed density field (Shandarin et al., 2006),
the method of discrete stochastic geometry, namely, Delaunay and
Voronoi tessellations (Schaap and van de Weygaert, 2000, Gaite,
2005). Grogin and Geller (1999, 2000) identified regions of low
galaxy density  by transformation of the point distribution of the
CfA2 survey into continuously defined number density field in
redshift space. We want to note also papers by Hoyle and Vogeley
(2002, 2004) who determined voids in PSCz, UZC and 2dFGRS surveys
by the voidfinder similar to (El-Ad and Piran, 1997) that divide
galaxy population on void-galaxies and wall-galaxies with
subsequent search for voids in distribution of wall-galaxies.
Original approach made by Croton et al. (2004) - some scaling
properties of 2dFGRS by the use of reduced VPF were revealed.

Nearly all of void-finders use some ad hoc parameter. For example
in (El-Ad and Piran, 1997) it defines in what tunnel between
galaxies void is allowed to penetrate and what tunnel is
suppressed. Our method in some sense is close to (El-Ad and Piran,
1997). It is simple, flexible and appropriate for our definition
of voids as regions completely free from galaxies.

In the volume of the sample under investigation an orthogonal
three-dimensional lattice is constructed so that to refer the
nodes of this grid to this or that void. The identification of
voids is made from larger to smaller ones. First, a
sphere with a largest radius of possible ones, which finds room
inside empty (free from galaxies) regions of considered volume
and geometric boundaries of the sample is detected. The voids
are supposed to be exactly inside the geometric boundaries
of the sample, therefore, the radius for a given node of
the grid is determined as the smallest of distances: from the
given node to nearest galaxy and a minimum distance from the
node to boundary of the sample.

Inside the volume of the sample (Fig.2) a consecutive search for
seed spheres (first the largest sphere is sought for) free from
galaxies executes with their subsequent expansion by means of
addition of spheres whose centers are located inside the already
fixed part of the void, and the radius $R_{sph}$ is no less than
that of the seed sphere multiplied by the coefficient $k=0.9$
(i.e. $R_{sph} > 0.9 \cdot R_{seed}$, where $R_{seed}$ is the
radius of the seed sphere). Thus, the spheres added to the void
intersect with the already referred to the void region by more
than 30\% of it's volume. Further an empty sphere is identified
again with the largest radius of possible with allowance made for
the void identified at the previous step of void-finding (all the
nodes of the grid referred to this void are excluded from the
consideration), then it is expanded etc. Finding of voids
continues until $R_{seed}$ is larger than a certain specified
threshold (in the present paper 0.5 Mpc).

When  $k=0.9$ the voids "represent" well enough the regions between
galaxies and at the same time preserve a sufficiently
regular form, which is convenient for subsequent approximation
of voids by ellipsoids. At $k=1$ voids will be strictly
spherical. When diminishing $k$, the voids begin to penetrate into
still smaller holes, and the shapes of voids become still more
irregular. At small $k$ one (first) void fills the greater
part of the sample volume. The value of the
coefficient $k=0.9$ used in this paper is compromising, chosen after
empirical detection of voids in distributions of points with
different properties. The voids thus constructed (with $k=0.9$) can be
detected in volumes of arbitrary shape. On the other hand,
the voids turn out to be separated from one another and ``thick''
enough throughout the entire their volume, which makes it
possible to approximate them by triaxial ellipsoids.

\renewcommand{\baselinestretch}{1.0}
\begin{table}
\caption{Local minivoids and bubbles}
\begin{tabular}{rrrrrrl}\\ \hline
Id    &
RA    &
DEC    &
$D_c$   &
$R_{seed}$  &
Volume &
Constellation   \\ \hline
      & h    &
deg    &
Mpc   &
Mpc  &
Mpc$^3$ & \\ \hline
 1    &19.02  &   2.9   &  3.80  &  3.33  & 443.6  & Aquila          \\
 2    & 3.63  & -12.6   &  4.40  &  2.94  & 159.8  & Eridanus        \\
 3    & 9.52  &  22.8   &  3.92  &  2.38  & 219.5  & Leo             \\
 4    & 8.51  & -40.0   &  5.46  &  2.06  &  61.2  & Vela            \\
 5    & 0.23  &  75.3   &  5.56  &  1.92  &  59.1  & Cepheus         \\
 6    & 0.82  & -86.0   &  5.39  &  1.86  &  85.2  & Octans          \\
\hline
 7    &23.77  & -34.3   &  5.99  &  1.53  &  27.6  &                 \\
 8    &13.70  &  30.0   &  6.09  &  1.43  &  19.8  &                 \\
 9    & 2.93  &  35.4   &  6.11  &  1.42  &  18.7  &                 \\
10    & 0.77  &  22.3   &  6.20  &  1.31  &  18.5  &                 \\
11    &11.44  &  50.0   &  6.18  &  1.27  &  24.2  &                 \\
12    &12.28  & -48.3   &  6.20  &  1.22  &  13.1  &                 \\
13    &15.50  &  60.6   &  6.36  &  1.16  &  13.6  &                 \\
14    &14.16  & -31.5   &  6.38  &  1.14  &  12.8  &                 \\
15    & 0.40  &  -8.1   &  6.37  &  1.13  &   9.3  &                 \\
16    & 6.69  &  -6.8   &  6.39  &  1.12  &   9.6  &                 \\
17    & 9.49  &  -9.5   &  6.40  &  1.12  &  14.3  &                 \\
18i   &10.96  & -54.8   &  2.24  &  1.10  &  11.1  &                 \\
19    & 5.18  &  19.6   &  6.41  &  1.09  &  11.7  &                 \\
20i   & 8.34  &  85.0   &  2.19  &  0.99  &  11.1  &                 \\
21    &15.91  & -52.5   &  6.50  &  0.99  &  12.9  &                 \\
22    &14.15  &  -0.8   &  6.55  &  0.95  &   8.4  &                 \\
23    & 2.09  & -47.5   &  6.54  &  0.94  &   5.0  &                 \\
24    &19.30  &  58.0   &  6.58  &  0.94  &   6.6  &                 \\
25    & 5.64  & -47.1   &  6.61  &  0.91  &   6.1  &                 \\
27    &23.27  &  41.4   &  6.62  &  0.90  &   5.4  &                 \\
29    &21.14  & -55.8   &  6.58  &  0.89  &   8.4  &                 \\
40    &22.34  & -22.9   &  6.72  &  0.70  &   7.0  &                 \\
42    &15.07  &  29.5   &  6.77  &  0.67  &   5.4  &                 \\
43i   &13.46  &  55.2   &  4.16  &  0.67  &   8.4  &               \\
\hline
\end{tabular}
\end{table}

\renewcommand{\baselinestretch}{2.0}

The detected voids are divided into voids lying completely
inside geometrical boundaries and into voids touching, when
constructed, the sample boundaries, whose volumes are, consequently,
underestimated, and the shapes are distorted by the
boundary effect.

In the present paper the step of the lattice $Step=0.1$ Mpc was used.
The volumes of the voids were estimated as $Step^3 \cdot N$, were
$N$ is the number of nodes of the grid inside the given void. The
described algorithm was applied to identification of voids
between galaxies populating the Local Volume. Since with growing
distance from the observer an abundance of galaxies
of low luminosities drops, and the number of galaxies with
accurately measured distances also decreases, we limited the
search of voids with a sphere of radius 7.5 Mpc around us.
This volume of 1766 Mpc$^3$ contains 355 galaxies about 90\% of which
distance estimates measured irrespective of their radial
velocities. Thus, the average volume for one galaxy is
$V_1 = 5.0$ Mpc$^3$ (the corresponding sphere radius is 1.06 Mpc).

The simulation of the procedure of voids detection at the Poisson distribution
of galaxies with the same number density as in our sample has shown that
voids with a volume of above 30 Mpc$^3$ have sufficiently high statistical
significance. There turned out to be six such voids in the
volume considered. These minivoids are listed in the upper part
of Table 1. Apart from them, our algorithm has found
another 24 empty volumes whose sizes exceed the average
volume $V_1$. These "bubbles" are enumerated in the
following 24 lines of Table 1. Finally, 52 smaller ``pores'' have been
detected by the algorithm. The majority of them may be artefacts.

\section{Census of the local minivoids and bubbles}

The columns of Table 1 present the following data on 6 minivoids and
24 bubbles: (1) --- ordinal number, (2,3) --- equatorial
coordinates of the center for the epoch J2000.0 in units of
hours and degrees; (4) --- distance to the center of the
void in Mpc; (5) --- radius of the seed sphere from which the
building up of the empty volume takes place; (6) --- volume
of a minivoid or a bubble in Mpc$^3$. It should be noted that the
majority of empty volumes that we have discovered come
in contact with the outer boundary of the sphere of radius 7.5 Mpc.
Only 3 volumes out of 30 are fully inside into
discussed sphere; they are marked by  ``i'' at the ordinal number.

The distribution of the detected empty volumes over the sky in
equatorial (a), galactic (b) and supergalactic (c) coordinates is
shown in Fig.3. Six largest minivoids are presented by big
circles, and 24 bubbles are depicted by small circles. All 355
galaxies of the Local Volume (filled circles) are broken up into
three classes according to distances, where the blue color
corresponds to the most nearby objects with $D<2.5$ Mpc, the green
color is for intermediate distances from 2.5 to 5.0 Mpc, while the
red color corresponds to the most distant ones.

 \begin{table}[hbtp]
\caption{Six the nearest bubbles/pores.}
\begin{tabular}{crrcrl}\\ \hline
Id    &  RA    & DEC     & $D_c$   & Volume &   Neighboring galaxies
  \\                \hline
      &   h    & deg     & Mpc   &  Mpc$^3$ &
  \\
\hline
72    & 7.53   & -6.7    & 0.83  &   1.5  & Phoenix, Fornax, LeoII, LMC
 \\
38    &12.60   & 18.8    & 1.23  &   2.1  & SexA, LeoI, LeoA, KK230
 \\
71    &23.81   &-48.8    & 1.32  &   1.1  & WLM, Tucana, E294-010
  \\
20    & 8.34   & 85.0    & 2.19  &  11.1  & M31, U8508, N1569, UA86, M82, KKH37
  \\
18    &10.96   &-54.8    & 2.24  &  11.1  & N3109, E321-014, I3104, Circinus
   \\
26    & 0.54   &-70.7    & 2.77  &   4.6  & I3104, I5152
   \\
\hline
\end{tabular}
\end{table}

The fourth part of the whole Local Volume is occupied by void $N 1$ in the
constellation Aquilla. It represents the front
part of the known Local void (Tully, 1988), which extends far beyond
the limits of the volume under consideration. At
the front boundary of this void a few dwarf galaxies (Sag DIG,
DD0210, NGC6822) adjoining the Local group are located.
As it follows from the distribution of voids in the galactic coordinates,
only a minor part of them ($\sim$1/4) can be
due to the Galactic extinction.

The distribution over the sky of the 24 bubbles with respect to one
another and to 6 largest minivoids does not show
considerable association. Such an effect could be expected in the case
where one or a few galaxies are situated inside
a large void, breaking it up into two neighboring volumes.

We present in Fig.4 a 3D map which demonstrates the shape and
mutual location of the 6 largest minivoids. To check an idea of
the presence/absence in the empty volumes of dwarf galaxies of
extremely low luminosity or intergalactic hydrogen clouds of low (
$M(HI)<10^5 M_{\odot}$) masses, one can use data on six most
nearby bubbles and pores (Table 2), the centers of which are
within 3 Mpc from us. The designations of columns here are the
same as in Table 1. The last column contains the names of galaxies
located at the boundaries of these small voids. The 3D map of the
distribution of the nearest bubbles/pores (Fig.5) shows tendency
to locating their centers along the plane of the Local
Supercluster.

\section{Shape and orientation of the minivoids and bubbles}

We approximate the surface of each empty volume identified by our
algorithm by a equivalent triaxial ellipsoid that has the same
moments of inertia as a body enclosed by the void. Their
distribution with respect to the minor-to-major axial ratio,
$c/a$, and the middle-to-major axial ratio, $b/a$, is shown in the
upper panel of Fig.6. The ratios of axes for 30 minivoids and
bubbles demonstrate a clear correlation characteristic of the
prolate but not oblate configurations. The minor and major axes
themselves are also correlated (lower panel of Fig.6), so that the
mean ellipticity remains almost independent of the size of a void.
Thus, the population of nearby voids looks like a set of oblong
potatoes (or haricots), among which there are no too flat or too
long shapes. The distribution of the directions of the major axes
of ellipsoids over the right half of the celestial hemisphere
(Fig.7) does not show noticeable anisotropy. However, the spatial
orientation of the major axes may take place in reality, with no
evidence in projection on the sky.

\section{Statistical properties of Local Volume}

Being arranged by the size of their volume, local voids follow the
Zipf power law $V(rank) \propto (rank)^{-z}$ with the exponent $z$
= 1.45 (upper panel of Fig.8), which can be treated as a fractal
character of distribution of voids (Gaite \& Manrubia, 2002).
Here, formally calculated fractal dimension is $D_z=3/z \simeq
2.07$. However, one can not correctly measure a fractal dimension
$D$ of 3D fractal set based on the Zipf law unless $2<D<3$ since
boundary of void has box dimension $D_b=2$ (Gaite 2006).
Application of our void-finder to different fractal sets showed
this clearly. The value of $D_z$ obtained by us is dangerously
close to 2, therefore, we may put only an upper constraint, $D
\leq 2.07$, on possible fractal dimension. We should note that we
use here 'fractal' language as a simple approximation to the data
and it will be interesting if other voidfinders would detect power
law in the Local Volume since physical description in the frame of
analytic model for the sizes of voids in galaxy distribution
(Furlanetto and Piran, 2005) predicts a peaked void distribution.
The total volume of the 6 minivoids is 58\% of the volume
considered. The addition of the 24 bubbles and 52 pores to them
increases the total volume of voids up to 75\% and 81\%,
respectively. The lower panel of Fig.7 shows a correlation
Gamma-function (Coleman \& Pietronero, 1992) of galaxies in the
Local Volume. The variation of density with scale $r$ in spheres,
$\Gamma^*(r)$, and spherical shells, $\Gamma(r)$, follows a power
law with an exponent $\gamma=1.48$ (lower panel of Fig.8), which
points to a formal fractal dimension $D_{\Gamma} = 3 - \gamma
\simeq 1.5$ of the distribution of the Local Volume galaxies. A
certain peculiarity in the $\Gamma$-function (bend) on a scale of
$\sim$3.5 Mpc may be associated with the characteristic distance
between the centers of groups in the Local Volume. Simple fractal
model of distribution of galaxies is not expected in modern
scenarios of hierarchical structure formation and moreover it was
rejected by statistical analysis (for example Tikhonov et al.,
2000) and appears to be more complex - multifractal (Jones et al.,
1992, McCauley, 2002, Jones et al., 2005). But nevertheless simple
fractal dimension gives essential geometric description of
distribution namely how galaxies fill given volume.

\section{Galaxies around minivoids and bubbles}

In different scenarios of formation of voids it is assumed
(Peebles, 2001, Hoffman et al., 1992) that galaxies in voids may
be different in their global parameters (luminosity, gas supply,
star formation rate) from the population of denser regions, in
particular, voids would be filled with galaxies of low luminosity.
To check this assumption, we have considered in the Local Volume
all galaxies which are located in layers 0.1 Mpc thick around
empty volumes. The distribution of these galaxies according to
their absolute blue magnitude $M_B$, hydrogen mass-to-luminosity
ratio $M(HI)/L_B$ and star formation rate per unit luminosity
$[SFR]/L_B$ in solar units are presented in three panels of Fig.9
(a, b, c, respectively). For comparison similar distributions of
the whole sample of galaxies in the same volume are shown by
opposite hatching. The data on hydrogen masses of galaxies have
been taken from the catalog CNG, and the measurements of the star
formation rate have been made from observations of the nearby
galaxies in the $H_{\alpha}$ line (Karachentsev \& Kaisin, 2006).
As it follows from these diagrams, the expected differences of
galaxies around the voids in the form of enhanced HI abundance and
higher star formation rate (Peebles, 2001) do have actually take
place. Their difference from the rest of galaxies manifests itself
at a significance level $(2-3)\sigma$. In order to check the
systematic differences of global properties of galaxies in
dependence on their environment, we have applied a distinct
approach similar to that of (Rojas et al., 2004). We have selected
galaxies within a radius of 0.8 Mpc around of which no neighbors
are present and compared them with the galaxies that have more
than 5 neighbors within 0.3 Mpc. The distribution of these
sub-samples in the parameters $M_B, M(HI)/L_B, [SFR]/L_B$ are
displayed in Fig.10 a, b, c, respectively. The differences noted
above for galaxies in loose and dense environment are seen here
even clearer.

It should be emphasized that the assumption that the regions of low
density are populated predominantly by objects of
low luminosity is not confirmed by observations. The differences
between galaxies around voids and in the general field are most
likely of opposite character as to their luminosity (Fig.8a and 9a).
The luminosity function of galaxies in dense environment looks
wider since close interactions and mergers of galaxies leading to
formation of faint ``tidal'' dwarfs and luminous mergers.

The galaxies surrounding nearby voids give us a unique opportunity
to check whether the expansion of empty volumes does occur.
In this case the galaxies located at the front boundary of a void
will have radial velocities on average lower than in the
homogeneous Hubble flow. On the farther side of a void the
galaxies, correspondingly, have excess of radial velocities.
We have selected on the near and back sides of the largest
mini-voids 27 galaxies with accurately measured distances and
velocities, and derived for them the mean difference of radial
velocities $<\Delta V> = (28\pm19)$ km s$^{-1}$. Here from the
radial velocity of each galaxy the regular Hubble component
$V = H_0 D$ was subtracted with $H_0$ = 72 km s$^{-1}$ Mpc$^{-1}$.
Thus, on the scale of effective radius of typical local minivoid
$\sim$1.5 Mpc an effect of local expansion with an amplitude
$\Delta H/H_0\sim(25\pm15)$\% is observed. The significance
of the expected effect could be increased by means of measuring
distances to other galaxies behind the nearby minivoids, which
is quite possible now with the facilities of the Hubble Space Telescope.

\section{Concluding remarks}

When identifying voids in the Local Volume, we were based on the assumption
that minivoids, bubbles and pores are completely free from galaxies.
This assumption, important in theoretical models (Peebles, 2001),
can be tested in subsequent observations. The "blind" survey of
the whole southern sky made in the HI line of neutral hydrogen
at the 64-m Parks radio telescope led to detection of new dwarf galaxies,
in particular, in the Zone of Avoidance of the Milky Way (Kilborn et al.,
2002, Staveley-Smith et al., 1998). A similar survey of the northern sky
is conducted in Jodral Bank (Boyce et al., 2001).
A much more sensitive blind survey in the HI line is now
conducting at the 300-m radio telescope in Arecibo
(Giovanelli et al., 2005). The minivoids Aquia, Leo, Eridanus
as well as six other bubbles of smaller size fall partially within
the zone of the Arecibo survey, DEC. = [0, +~38$^{\circ}$]. We
expect that the results of this survey will have a chance to confirm
or reject the idea of complete absence of galaxies in nearby voids.
The detection of new nearby galaxies and measurement
of their distances from the luminosity of red giant branch
stars will allow a more detailed study of the shape and kinematics
of the nearest voids. As was noted above, the Local void in
Aquila, detected by Tully, extends far beyond the limits
of the Local Volume. By the present time, the number of nearby
galaxies with radial velocities $< 3000$ km s$^{-1}$ has reached
about seven thousand. This new observational material presents a
possibility of detecting and investigating voids by two orders
of magnitude exceeding the volume considered here.

\acknowledgements{The authors are grateful to S. Shandarin, D. Makarov,
J. Gaite and B. Tully for useful advices and discussions. Support
associated with HST programs 9771 and 10235 was provided by NASA
through a grant from the Space Telescope Science Institute,
which is operated by the Association of Universities for
Research in Astronomy, Inc., under NASA contract NAS5--26555.
This work was also supported by RFFI grant 04--02--16115
and by grant Rosobrazovanie RNP 2.1.1.2852.}

{}
\newpage
\begin{figure}
\centerline{
\includegraphics[]{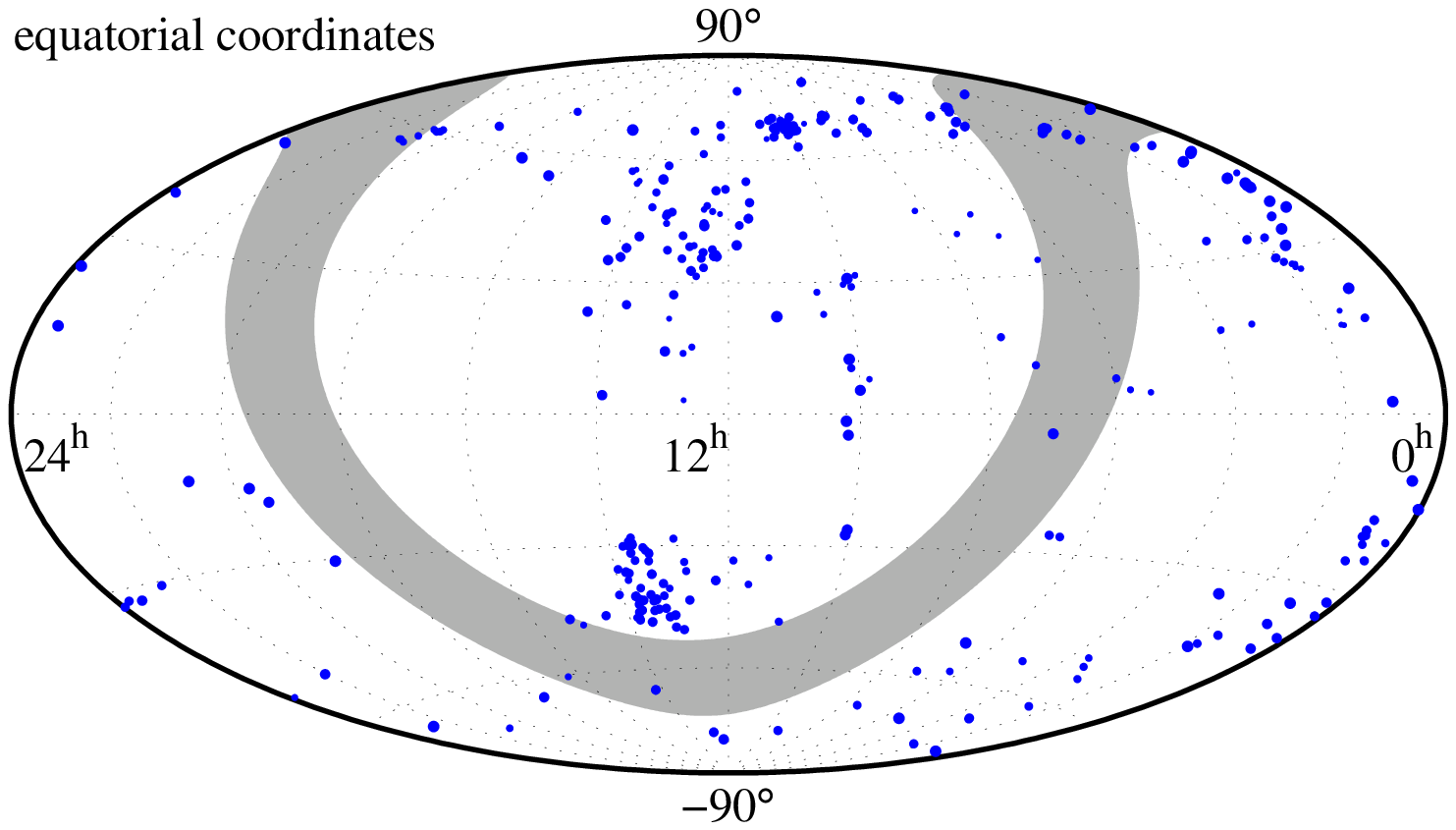}
} \figcaption{All-sky distribution of Local Volume galaxies with
distances less than 7.5 Mpc in equatorial coordinates. Sizes of
circles decrease in accordance with increasing of distances of
galaxies. The shaded area marks the Zone of Avoidance in the Milky
Way.}
\end{figure}

\begin{figure}
\centerline{
\includegraphics[]{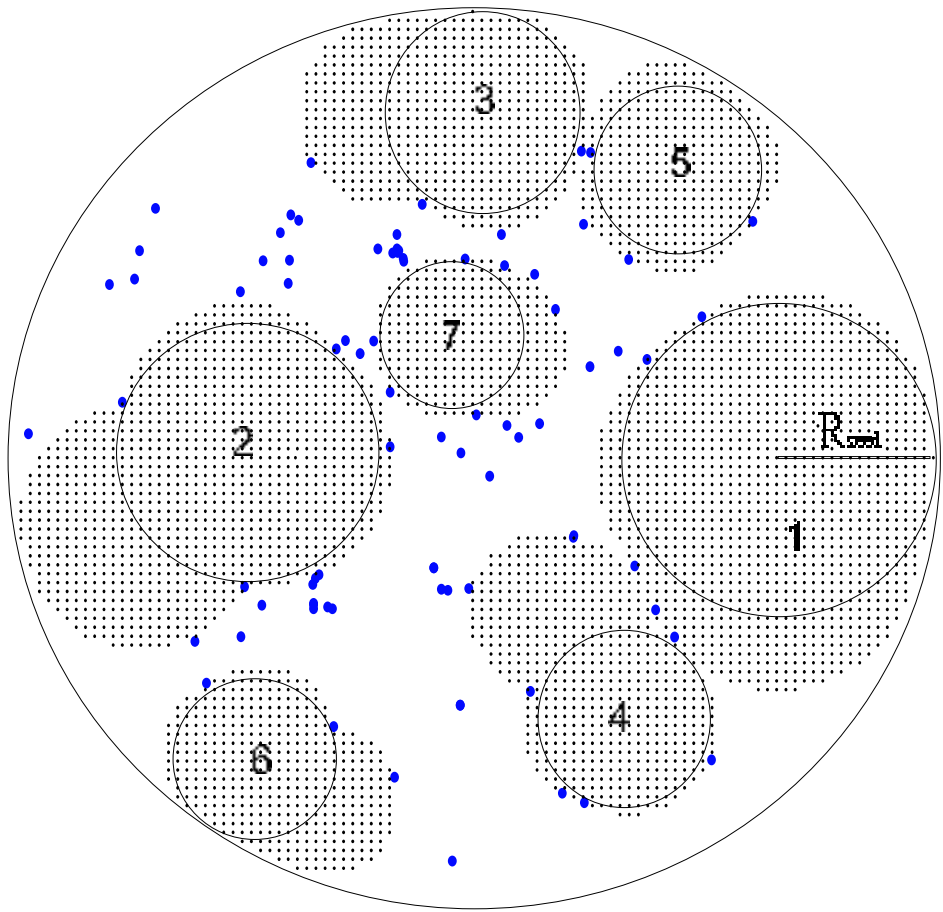}
}
\figcaption{Illustration of action of the void-finding algorithm
 in a case of 2D point-like distribution. Seed spheres and voids growing
from them are shown. The numerals indicate the order of detection of voids
from large seed spheres to smaller ones.}
\end{figure}

\begin{figure}[p]
\centerline{
\includegraphics[scale=.7]{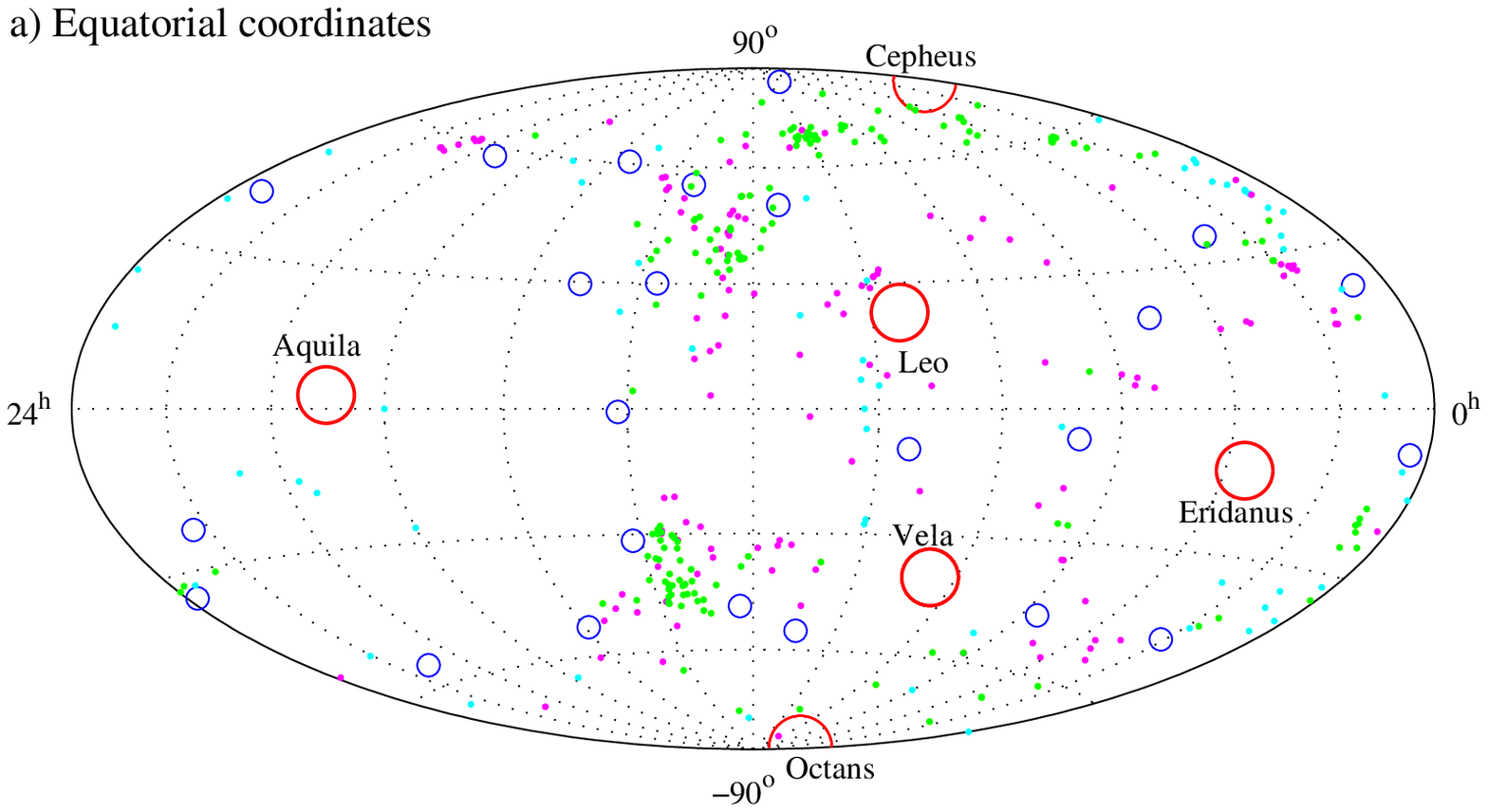}
}
\centerline{
\includegraphics[scale=.7]{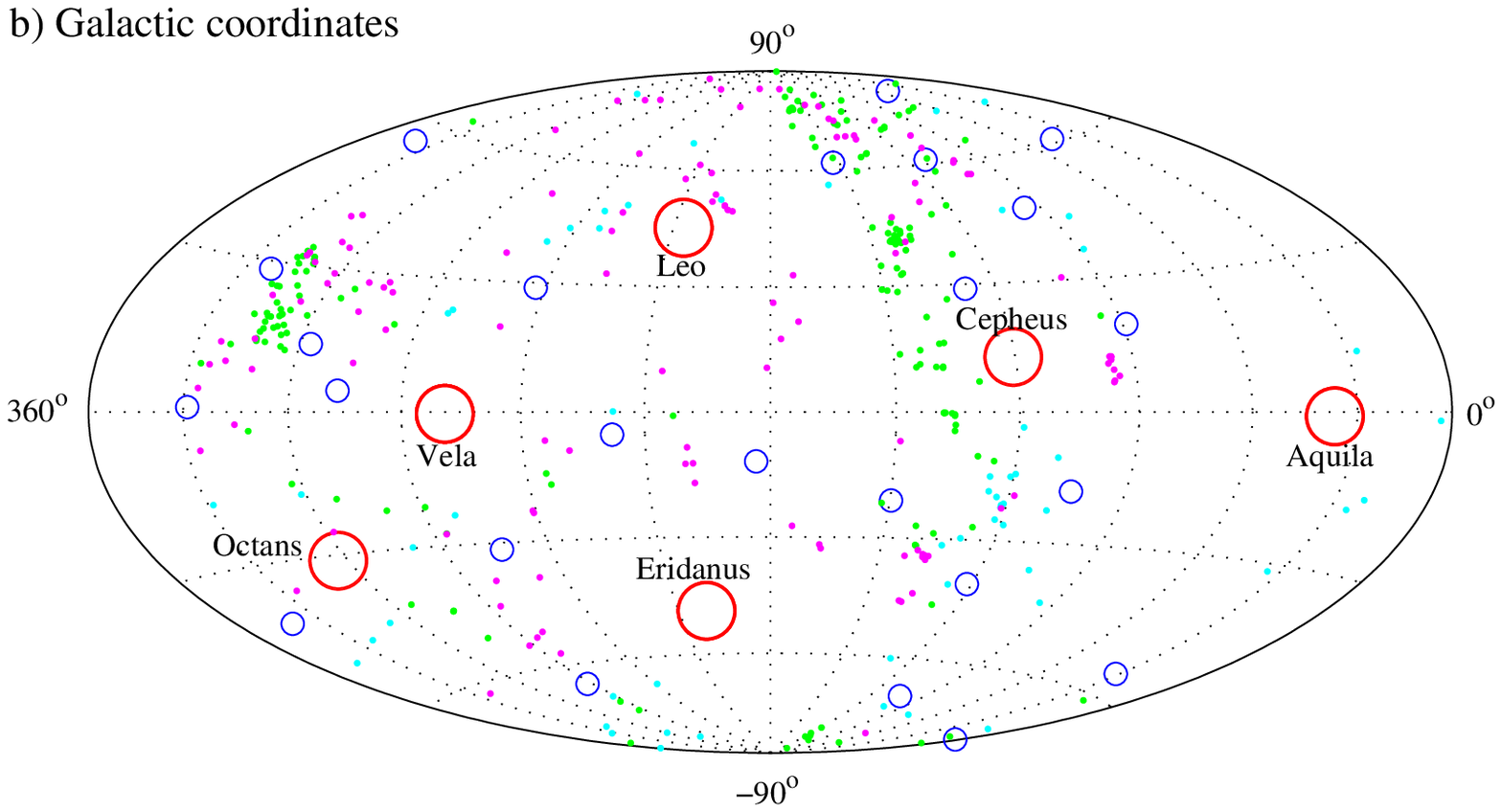}
}
\centerline{
\includegraphics[scale=.7]{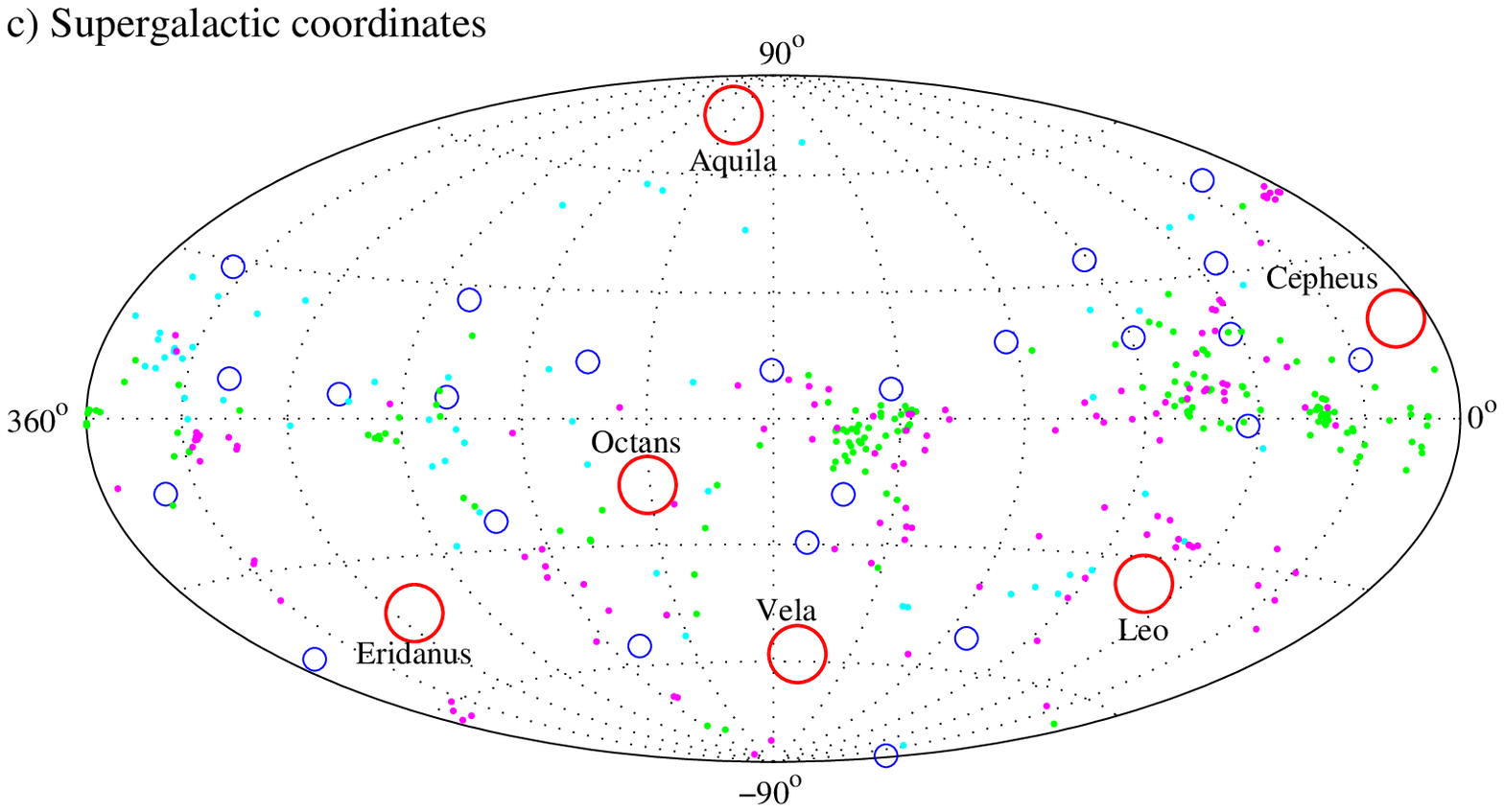}
}
\figcaption{Distribution over the celestial sphere of
 six large minivoids (large red circles around void centers),
 24 bubbles (small blue circles) and the Local Volume galaxies
 with distances $D < 2.5$ (blue), $2.5 < D < 5$ (green)
 and $5 < D < 7.5$ (red).}
\end{figure}

\topmargin=-1cm
 \newpage
\begin{figure}[p]
\centerline{
\includegraphics[]{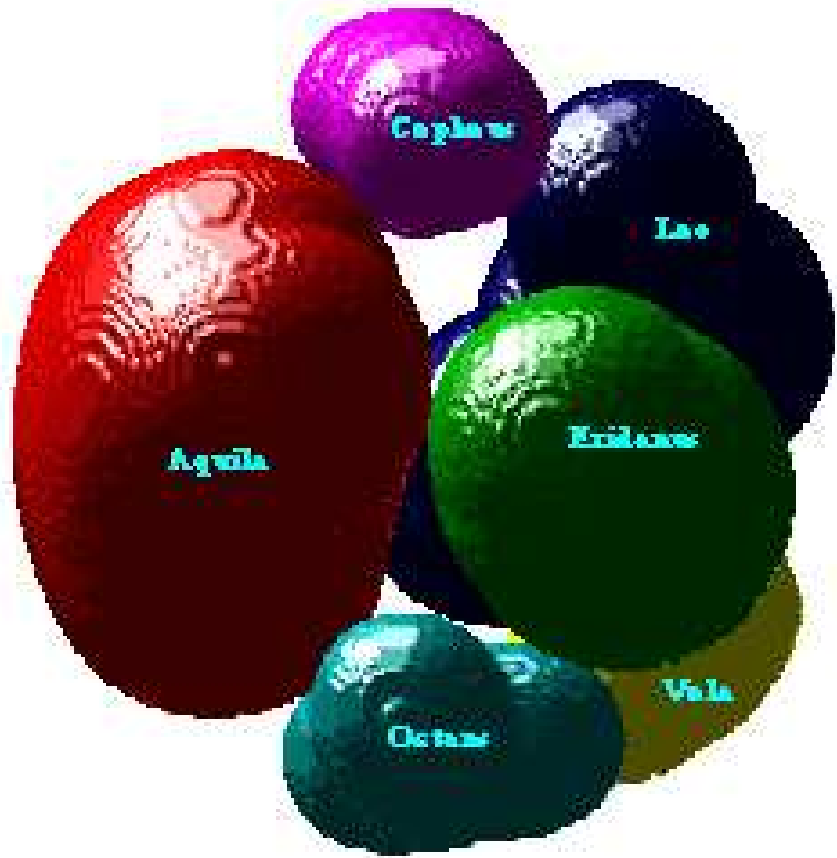}
}
\figcaption{Distribution of six large minivoids within the sphere of
 radius 7.5 Mpc.}
\end{figure}

\begin{figure}[p]
\centerline{
\includegraphics[]{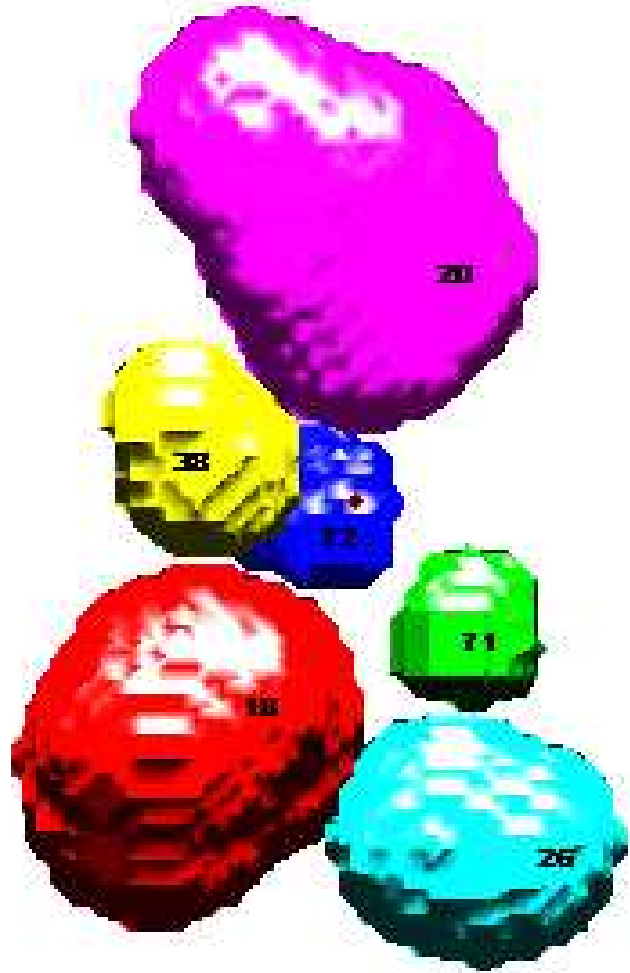}
}
\figcaption{Distribution of two nearest bubbles (No 20 and 18) and
 four pores (No 72, 38, 71, 26) with distances to their centers less
 than 3 Mpc. Perspective is XZ plane in equatorial coordinates.
 The circle indicates the Milky Way location.}
\end{figure}

\newpage
\begin{figure}[p]
\centerline{
\includegraphics[angle=0,scale=.9]{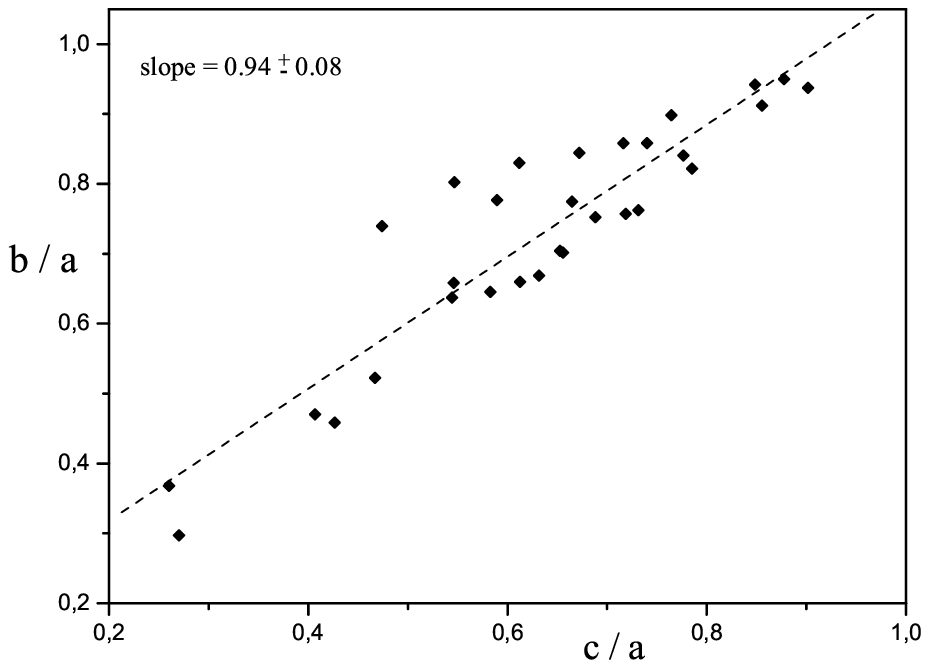}
}
\centerline{
\includegraphics[angle=0,scale=.9]{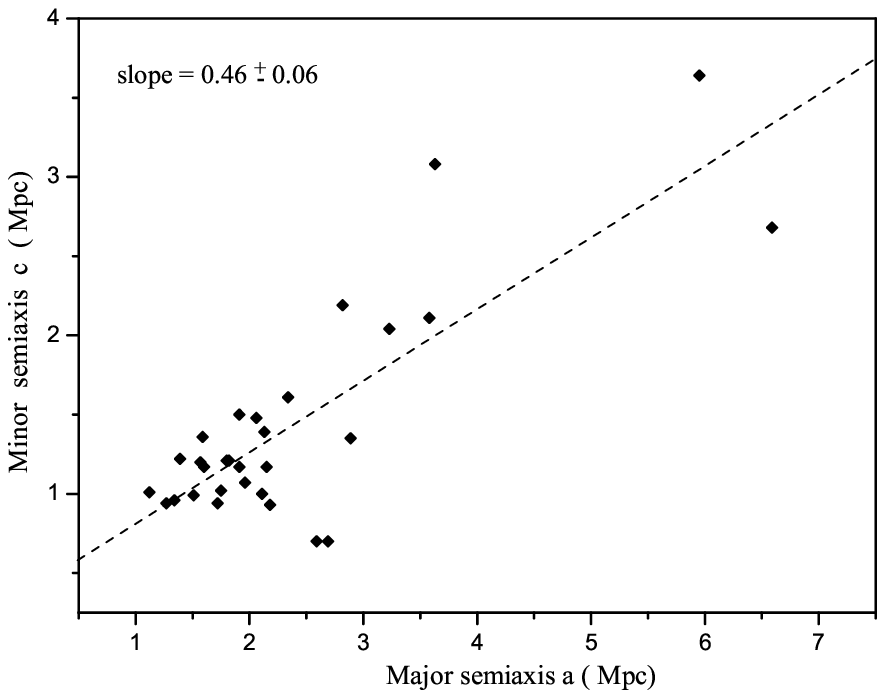}
} \figcaption{(Top)  Relationship among axial ratios of 30
minivoids and bubbles:
 minor-to-major, c/a, and middle-to-major, b/a, axial ratios of equivalent
 ellipsoids.
(Bottom)  Relationship among the semi-major and semi-minor axes of equivalent
ellipsoids for the same voids.}
\end{figure}

\begin{figure}[p]
\centerline{
\includegraphics[]{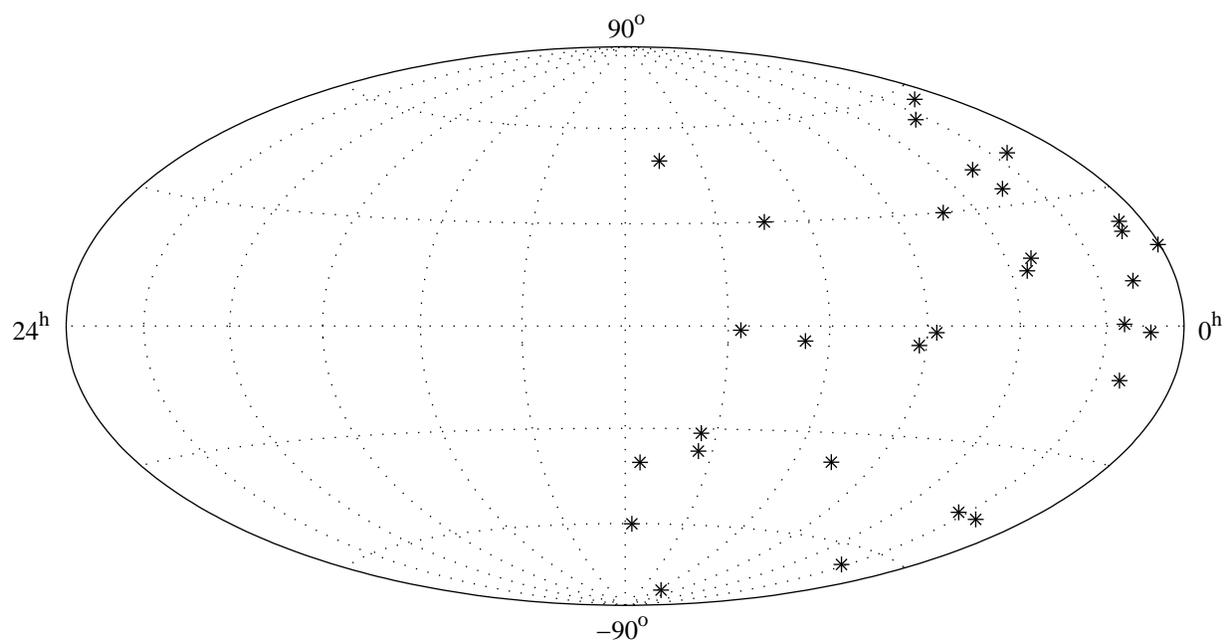}
}
\figcaption{A view of directions of the major axes of 30 voids
 distributed over the right half of the celestial sphere.}
\end{figure}

\begin{figure}[p]
\centerline{
\includegraphics[angle=0,scale=.9]{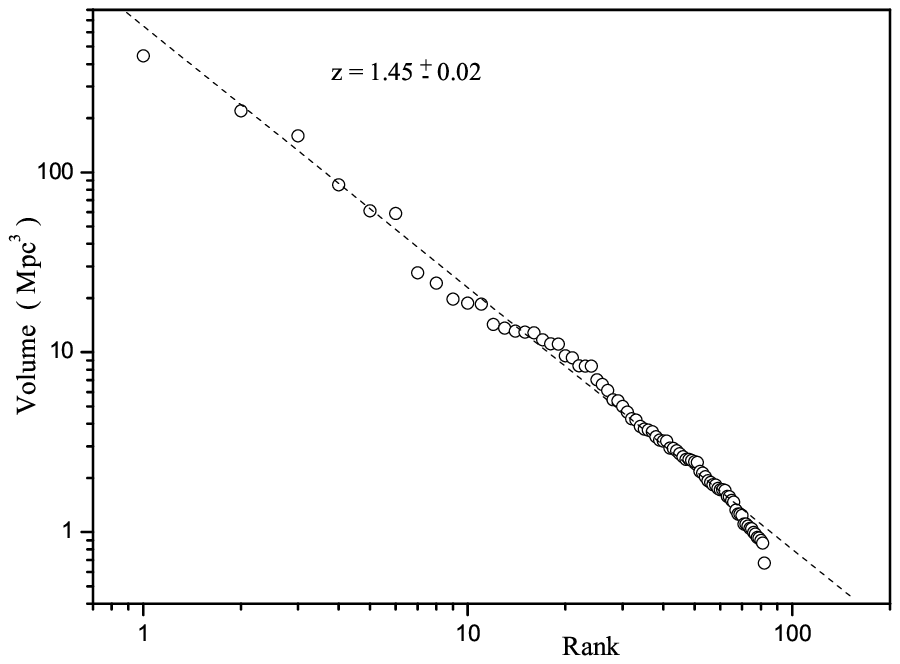}
}
\centerline{
\includegraphics[angle=0,scale=.9]{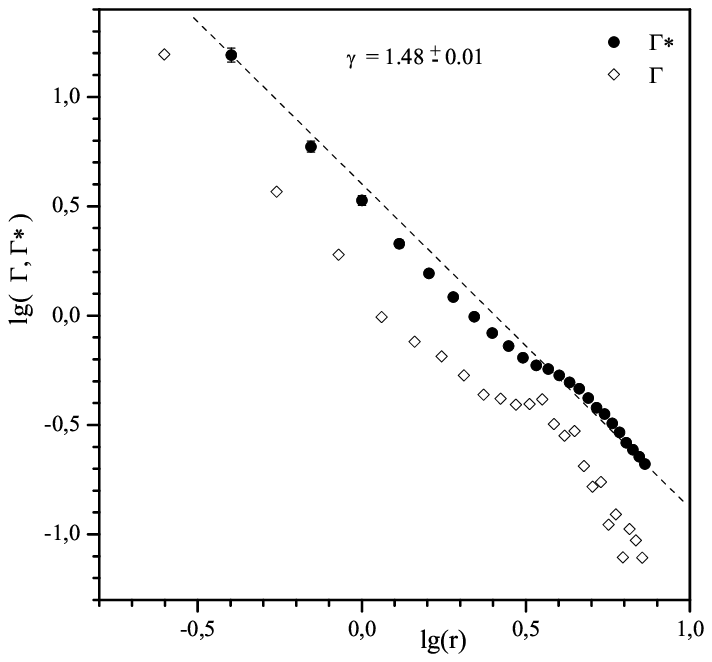}
}
\figcaption{(Top) The void volumes versus their ranks in
   $log_{10}$ scale. Power law is evident.
(Bottom)  Correlation Gamma-function of 355 galaxies in Local Volume within
  a sphere of radius 7.5 Mpc.}
\end{figure}

\begin{figure}[p]
\centerline{
\includegraphics[angle=0,scale=.6]{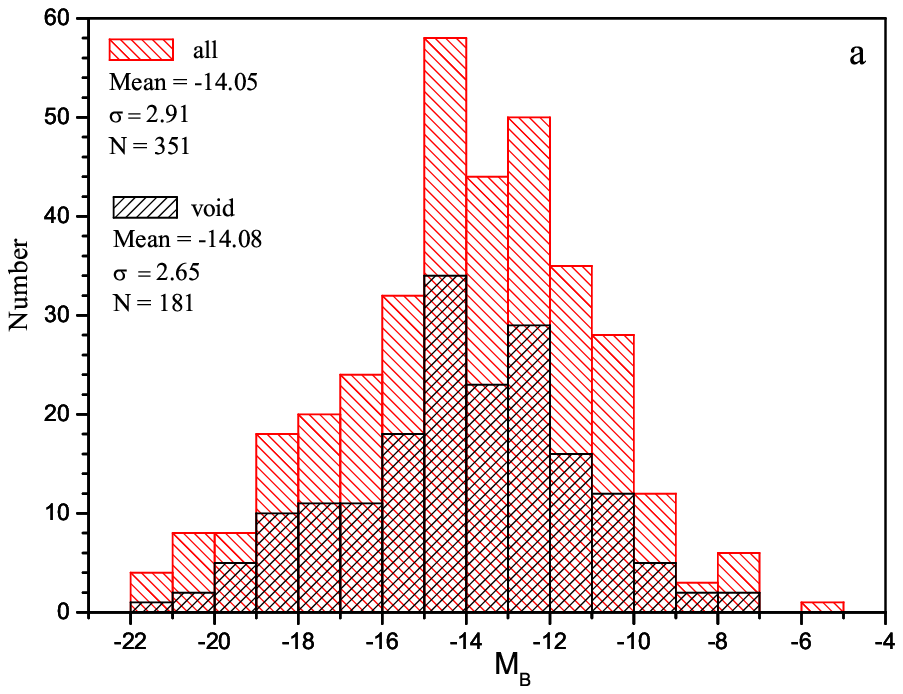}
}
\centerline{
\includegraphics[angle=0,scale=.6]{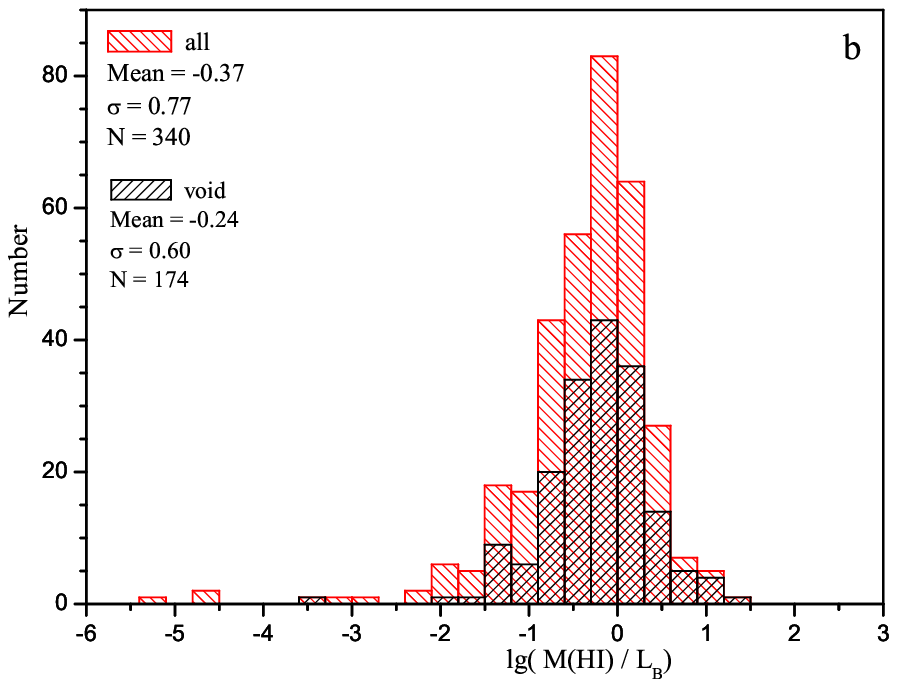}
}
\centerline{
\includegraphics[angle=0,scale=.6]{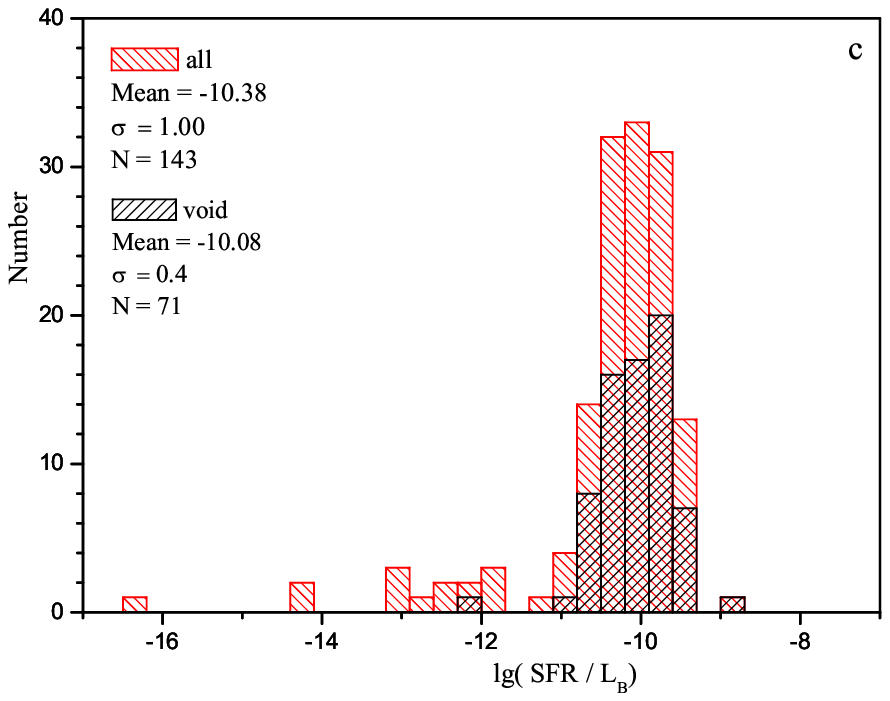}
}
\figcaption{(a) - B-band absolute magnitude distribution of
 the Local Volume galaxies and galaxies around local voids;
 (b) - ratio of hydrogen mass to blue luminosity in solar units;
 (c) - star formation rate per unit of blue luminosity.}
\end{figure}

\begin{figure}[p]
\centerline{
\includegraphics[angle=0,scale=.6]{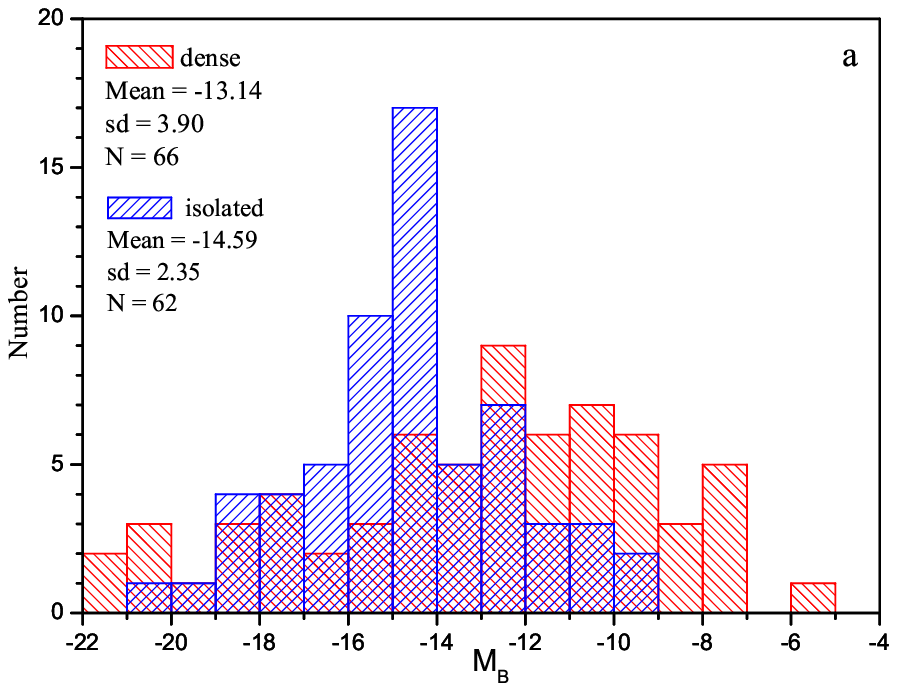}
}
\centerline{
\includegraphics[angle=0,scale=.6]{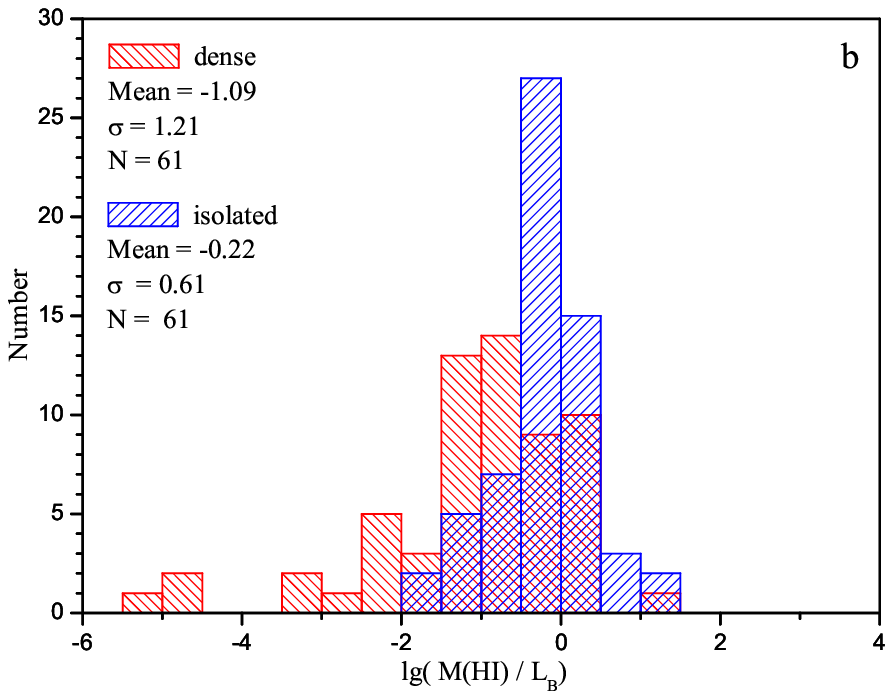}
}
\centerline{
\includegraphics[angle=0,scale=.6]{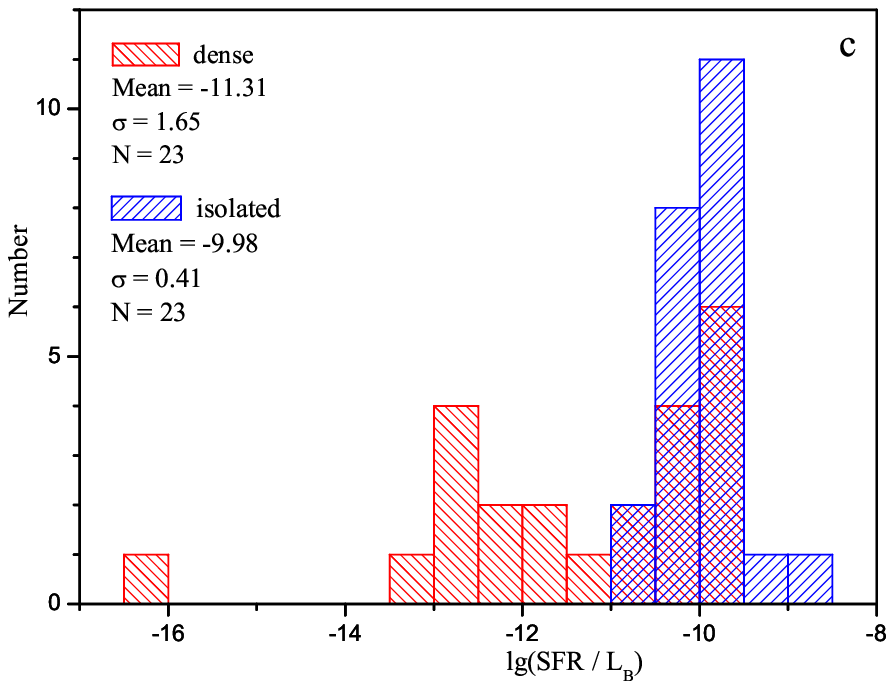}
}
\figcaption{(a) - B-band absolute magnitude distribution of
isolated galaxies and galaxies in dense environment of the Local
Volume; (b) - ratio of hydrogen mass to blue luminosity in solar
 units; (c) - star formation rate per unit of blue luminosity.}
\end{figure}


\begin{thebibliography}{}

\bibitem[]{472} Aikio J., Mahonen P., 1998, ApJ, 497, 534
\bibitem[]{473} Boyce P.J., et al. 2001, ApJ, 560, L127
\bibitem[]{474} Colberg J.M., Sheth R.K., Diaferio A. et al., 2005, MNRAS, 360, 216
\bibitem[]{475} Coleman P.H., Pietronero L., Physics Reports, 1992, 213, 311
\bibitem[]{477} Croton D., Colles M., Gaztanaga E. et al., 2004, MNRAS, 352, 828
\bibitem[]{478} de Lapparent V., Geller M.J., Huchra J.P., 1986, ApJ, 302, L1
\bibitem[]{479} Einasto J., Einasto M., Gramann M., 1989, MNRAS, 238, 155
\bibitem[]{480} El-Ad H., Piran T., 1997, ApJ, 491, 421
\bibitem[]{481} Furlanetto S. and Piran T., 2005,
\bibitem[]{481} Gaite J., Manrubia S.C., 2002, MNRAS, 335, 977
\bibitem[]{482} Gaite J., 2005, Eur.Phys.J. B47, 93, astro-ph/0506543
\bibitem[]{483} Gaite J., 2006, preprint (astro-ph/0603572)
\bibitem[]{484} Ghigna S., Borgani S., Bonometto S., et al. 1994, ApJ, 437, L71
\bibitem[]{485} Giovanelli R., Haynes M., Karachentsev I. et al., 2005, AJ, 130, 2598
\bibitem[]{486} Gottlober S., Lokas E., Klypin A., Hoffman Y., 2003, MNRAS, 344, 715
\bibitem[]{487} Gregory S.A., Thompson L.A., 1978, ApJ, 222, 784
\bibitem[]{488} Grogin and Geller, 1999, AJ, 118, 2561
\bibitem[]{489} Grogin and Geller ,2000, AJ, 119, 32
\bibitem[]{490} Hoffman Y., Silk J., Wyse R., 1992, ApJ, 388, 13
\bibitem[]{491} Hoyle F. and Vogeley M., 2002, ApJ, 566, 641
\bibitem[]{492} Hoyle F. and Vogeley M., 2004, ApJ, 607, 751
\bibitem[]{493} Icke V., 1984, MNRAS, 206, Short Comunication, 1P
\bibitem[]{494} Icke V., Van de Weygaert R., 1991, QJRAS, 32, 85
\bibitem[]{495} Iwata I., Ohta K., Nakanishi K., et al., 2004, in "Nearby
   Large-Scale Structures and the Zone of Avoidance", ASP Conference
   Series, v. 329, eds. A.P.Fairall, P.A.Woudt, p. 59
\bibitem[]{496} Joeveer M., Einasto J., Tago E., 1978, MNRAS, 185, 357
\bibitem[]{497} Jones B., Coles P., Martinez V., 1992, MNRAS, 259, 146
\bibitem[]{497} Jones B., Martinez V., Saar E., Trimble V., 2005, Reviews of Modern Physics, 76, 1211
\bibitem[]{498} Karachentsev I.D., 1994, Astron. \& Astrophys. Trans. 6, 1
\bibitem[]{499} Karachentsev I.D., Kaisin S.S., 2006, in preparation
\bibitem[]{500} Karachentsev I.D., Karachentseva V.E., Huchtmeier W.K., Makarov D.I.,
     2004, AJ, 127, 2031 (=CNG)
\bibitem[]{501} Karachentseva V.E., Karachentsev I.D., Richter G.M., 1998, A\&A, 135, 221
\bibitem[]{502} Kauffmann G., Fairall A.P., 1991, MNRAS, 248, 313
\bibitem[]{503} Kilborn, V.A., Webster, R.L., Staveley-Smith, L. et al. 2002, AJ, 124, 690
\bibitem[]{504} Kirshner R.P., Oemler A., Schechter P.L., Shectman S.A., 1981, ApJ, 248, L57
\bibitem[]{505} McCauley J.L., Physica A, 2002, 309, 183, astro-ph/9703046
\bibitem[]{506} Patiri S.G., Betancort-Rijo J., Prada F., et al., 2005, astro-ph/0506668
\bibitem[]{507} Peebles P.J.E., 2001, ApJ, 557, 495
\bibitem[]{508} Plionis M., Basilakos S., 2002, MNRAS, 330, 399
\bibitem[]{509} Regoes E., Geller M., 1991, ApJ, 377, 14
\bibitem[]{510} Rojas R., Vogeley M., Hoyle F., Brinkmann K., 2004, ApJ, 617, 50
\bibitem[]{511} Rood H.J., Annual review of astronomy and astrophysics, 1988, 26, 245
\bibitem[]{512} Ryden B.S., Melott A.L., 1996, ApJ, 470, 160
\bibitem[]{513} Sahni V., Sathiaprakash B., Shandarin S., 1994, ApJ, 431, 20
\bibitem[]{514} Schaap, van de Weygaert, 2000, A\&A, 363, L29-L32
\bibitem[]{515} Shandarin S., Feldman H.A., Heitmann K., Habib S., 2006, MNRAS, 367, 1629
\bibitem[]{516} Sheth R.K., Van de Weygaert R., 2004, MNRAS, 350, 517
\bibitem[]{517} Staveley-Smith, L., Juraszek, S., Koribalski, B.S. et al. 1998, AJ, 116, 2717
\bibitem[]{518} Tikhonov A.V., Makarov D.I., Kopylov A.I., 2000, Bull. SAO RAS, 50, 39, astro-ph/0106276
\bibitem[]{519} Tully R.B., 1988, Nearby Galaxy Catalog, Cambridge University Press
\bibitem[]{520} Van de Weygaert R., 1994, A\&A, 283, 361
\bibitem[]{521} Van de Weygaert R., van Kampen E., 1993, MNRAS, 263, 481
\bibitem[]{522} Zeldovich, Ia. B., Einasto, J., Shandarin, S. F., 1982, Nature, 300, 407
\end{thebibliography}
\end{document}